\documentclass[sigchi]{acmart}
\vbadness=\maxdimen
\PassOptionsToPackage{hyphens}{url}
\usepackage{booktabs} 

\usepackage{xcolor}

\usepackage{balance}

\setcopyright{none}

\acmDOI{10.475/123_4}

\acmISBN{123-4567-24-567/08/06}

\acmConference[CHI '19]{}
\acmYear{2019}
\copyrightyear{2019}

\acmPrice{}

\begin{document}
\title{Analyzing the Use of Camera Glasses in the Wild}

\hyphenation{tech-no-lo-gy}
\hyphenation{post-ad-opt-ion}

\settopmatter{printacmref=true, authorsperrow=4}

\author{Taryn Bipat}
\email{tbipat@uw.edu}
\affiliation{%
  \institution{University of Washington, Snap Inc.}
  }

\author{Maarten Willem Bos}
\email{mbos@snap.com}
\affiliation{%
  \institution{Snap Inc.}
}

\author{Rajan Vaish}
\email{rvaish@snap.com}
\affiliation{%
  \institution{Snap Inc.}
}

\author{Andr\'es Monroy-Hern\'andez}
\email{amh@snap.com}
\affiliation{%
  \institution{Snap Inc.}
 }


\begin{abstract}
Camera glasses enable people to capture point-of-view vi\-de\-os using a common accessory, hands-free. In this paper, we investigate how, when, and why people used one such product: Spectacles. We conducted 39 semi-structured interviews and surveys with 191 owners of Spectacles. 
We found that the form factor elicits sustained usage behaviors, and opens opportunities for new use-cases and types of content captured. We provide a usage typology, and highlight societal and individual factors that influence the classification of behaviors. 
\end{abstract}

%
%


\copyrightyear{2019} 
\acmYear{2019} 
\setcopyright{acmcopyright}
\acmConference[CHI 2019]{CHI Conference on Human Factors in Computing Systems Proceedings}{May 4--9, 2019}{Glasgow, Scotland UK}
\acmBooktitle{CHI Conference on Human Factors in Computing Systems Proceedings (CHI 2019), May 4--9, 2019, Glasgow, Scotland UK}
\acmPrice{15.00}
\acmDOI{10.1145/3290605.3300651}
\acmISBN{978-1-4503-5970-2/19/05}


 \maketitle

\section{Introduction}

The range of form factors for wearable devices has expanded dramatically in recent years, from watches to clothing, helmets, and more. As people expand their technology use from mobile phones to wearables, new research questions emerge about how, why, and where these technologies are used. In this paper we focus on studying \textit{camera glasses}, a type of smart glasses that have the camera as their core functionality.

Researchers have classified the functionality of, and tried to predict the motivations for early adoption of, various types of wearables \cite{park2015wearables,bhat2014wearable}. However, there is not much work on how these devices are actually used once they are adopted. Furthermore, when it comes to camera glasses, most of the existing research focuses on experimental applications \cite{McAtamney:2006:EEW:1124772.1124780,Byrne:2008:STT:1531826.1531832,Muralidhar:2016:DWF:3012709.3012733} and not much on how everyday people use camera glasses in their life, outside research settings.

In this work we seek to further understand the postadoption use and social implications of smart glasses. We do this by studying the everyday use of Spectacles, a set of commercially available camera glasses that its manufacturer describes as ``sunglasses that capture your world'' and a ``hands-free camera'' \cite{spectacles}. Spectacles' users capture 10-second videos by pressing a button at the top left corner of the device. Pressing the button again adds 10 seconds to the recording, for a maximum of 30 seconds. Pressing-and-holding captures a still photo.

The content is automatically and wirelessly transferred from the device's storage to the user's Snapchat application on their smartphone, which people can use to view and share their content. The videos and photos are recorded in a high-definition (HD) circular shape, but they are exportable in circular, square, or wide-screen formats. 

We see ``camera glasses,'' like Spectacles, as part of a ``smart glasses'' category of wearables that  include VR/AR headsets and glasses with speakers \cite{Vue}. 
Unlike other smart glasses, camera glasses consist only of capturing technology. To make the distinction clear, we will refer to Spectacles as camera glasses.

We conducted 39 semi-structured interviews and surveys with 191 owners of Spectacles. We found that participants used their camera glasses primarily during social interactions and while being physically active. Furthermore, participants captured their own perspective for preserving their memories and to share them with others. For them, camera glasses supplement or even replace other capturing devices. Users' individual lifestyles---from their job to the events they attend---dictate how they use camera glasses. Lastly, we found that individual and societal boundaries affect how people use camera glasses. People take societal views of camera glasses into account when deciding in what scenarios to use them.

\section{Literature Review}
Some of the earliest descriptions of wearables focused on three key modes of interactions they enable: \textit{constancy}, always ready to be used; \textit{augmentation}, enhancing analog tasks; \textit{mediation}, filtering incoming information and protecting users' privacy \cite{Mann98}. 

Recently this model was expanded \cite{wright2014wearable}, adding a focus on form factors. The updated model describes wearables as electronics and computers that are integrated into clothing and other accessories worn comfortably on the body. This recent work argues that wearables have the potential to improve people's health by becoming integrated into people's way of life.

Despite this interest in how wearables fit in people's everyday lives, a lot of the research on smart glasses has focused on experimental hardware and niche applications\cite{Zhang:2014:SIV:2639108.2639119,Rallapalli:2014:EPA:2639108.2639126,le2013eye,Xiong:2017:ISG:3081333.3081343}. Even when investigating commercially-available products like Google Glass, researchers focused on niche medical applications \cite{Malu:2014:OGP:2661334.2661400}. 

Another line of research on smart glasses has centered around identifying potential early adopters, and their motivations \cite{Rauschnabel:2015:BSG:2778387.2778668,rauschnabel2016augmented}. This work is elucidating but it does not get at what people actually do once they purchase these devices. For example, Kalatari and colleagues \cite{kalantari2017consumers} highlighted the lack of qualitative studies on the types of usage people exhibit \textit{after} acquiring smart glasses. 

Recent research has also found that wearables' adoption, although increasing, has been slower and more varied than initially expected \cite{kalantari2017consumers}. This also highlights the importance of understanding sustained use ``in the wild'' when it does happen.

A study like ours is needed to delve deeper into sustained behaviors; revealing the need to understand concerns around privacy, fashion trends, technological challenges and more \cite{kalantari2017consumers}.

\subsection{Social Implications of Wearables}
Prior research has shown that wearables' style and the fact that they are worn close to the body contribute to the technology being perceived as a companion and fashion item, rather than just as a tool \cite{Kao:2017:EIP:3064663.3064686}. This has perhaps influenced the design of camera glasses, many of which are built to look like regular sunglasses. 

Research has also shown that body-worn cameras can create tensions in certain social settings \cite{Koelle:2017:YHU:3098279.3122123} due to privacy concerns. 
It is worth noting that Spectacles, the camera glasses described in this paper, mitigate these challenges through the use of LEDs that inform bystanders that recordings are occurring.  
We identified the opportunity to contribute to the literature by studying how camera glasses are used post-adoption. As part of this, we frame our investigation around two dimensions: \emph{activities} and \emph{time}. We build on prior work on ``context'' by Dey and colleagues who identified four contextual dimensions for technology: location, identity, activities, and time \cite{abowd1999towards}.

Although there is not much research specifically on Spectacles, it is worth noting that prior work on smart glasses referred to Spectacles as a tool for lifelogging \cite{chow2017neurocognitive,Koelle:2017:YHU:3098279.3122123,fan2018deepdiary} and storytelling  \cite{bossert2017innovative}, a point-of-view camera platform \cite{Chong:2017:DGT:3139486.3131902}, and as an extension of the Snapchat mobile app \cite{khan2017snapchat,Lanigan:2017:SGL:3154054}.

\section{Methods}
Since we aimed to understand not only how and where, but also why people use camera glasses, we decided to use both surveys and interviews. The semi-structured interviews allowed us to understand who the users are and their lived experiences with camera glasses. For detailed information, both protocols can be found in the supplementary documents.

We partnered with the Spectacles team at Snap Inc., the company that created Spectacles, to reach out to people who had opted to be contacted and had used their camera glasses for at least two weeks. The research protocol was reviewed by an internal team of privacy experts at Snap Inc. before being run. 

We do not differentiate between the two versions\footnote{one released in November, 2016, and the other in April, 2018} of Spectacles that were available at the time of this study because their basic functionality---as it relates to this paper---remained largely the same. That said, it is worth noting that because the second version is water-resistant, some of the use cases we describe might vary slightly across versions.

\subsection{Surveys}
The 191 participants who answered our survey were recruited via email approximately two weeks after they had successfully paired their camera glasses with a phone for the first time. We emailed participants during May and June of 2018. The survey included multiple-choice and open-ended questions about who participants were, how they used their device, and their reasons for using it. Through these surveys, we were able to understand users' initial perceptions of the device, reasons for purchasing it, and how it could be improved. The open-ended responses were hand-coded, and then bucketed into unique categories or placed within similar multiple-choice buckets.

\subsection{Semi-structured interviews}
We used semi-structured interviews to obtain qualitative data of people's experiences. We recruited 39 participants who had owned their camera glasses for three to twenty months. Participants stemmed from seven different countries: USA, UK, The Netherlands, Sweden, Israel, Canada, and France. Participants included students, scientists, marketing specialists, entertainers, car mechanics, journalists, and IT professionals.

Each video interview lasted approximately 45 minutes and was conducted over video conference, which allowed us to broaden our geographic reach. All interviews were recorded with participants' consent, and the interviewer took detailed notes of the conversations. Participants received a USD \$50 gift card at the end of the session.

\subsection{Data Analysis}
To analyze the notes taken during the interviews, the first author developed a common set of themes that emerged from our data through thematic analysis \cite{braun2006using}. Two researchers met on a weekly basis to extract and refine the coding scheme. After all user interviews were conducted, the main findings from each of the participants were listed and then reviewed by all authors of the paper. Subsequently, we used the interview recordings to identify additional content related to those themes.

\section{Results}
\begin{figure}
\centering
	\includegraphics[width=\linewidth] {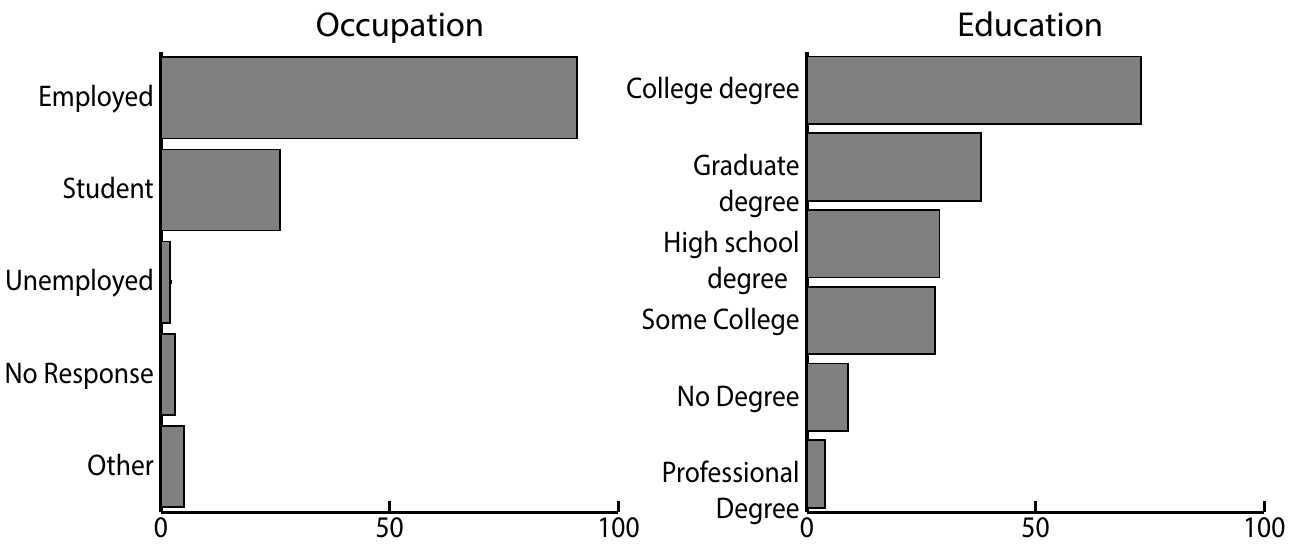}
	\caption{Survey Participants' Occupation and Education.}
     ~\label{fig:demo} 
\end{figure}

Most participants were employed and highly educated, or in the process of getting their education (see Figure \ref{fig:demo}). This might correspond to the state of adoption for wearables in general, as early adopters tend to skew towards higher socio-economic status. 

We structure the results in terms of when (time), how (scenarios), and why (activities) people use their camera glasses. We intertwine the results from the surveys and the interviews as they displayed the same patterns, with the interviews providing more in-depth insights to the survey results. 

\subsection{Time: When are camera glasses used?}
\begin{figure} 
	\includegraphics[width=\linewidth]{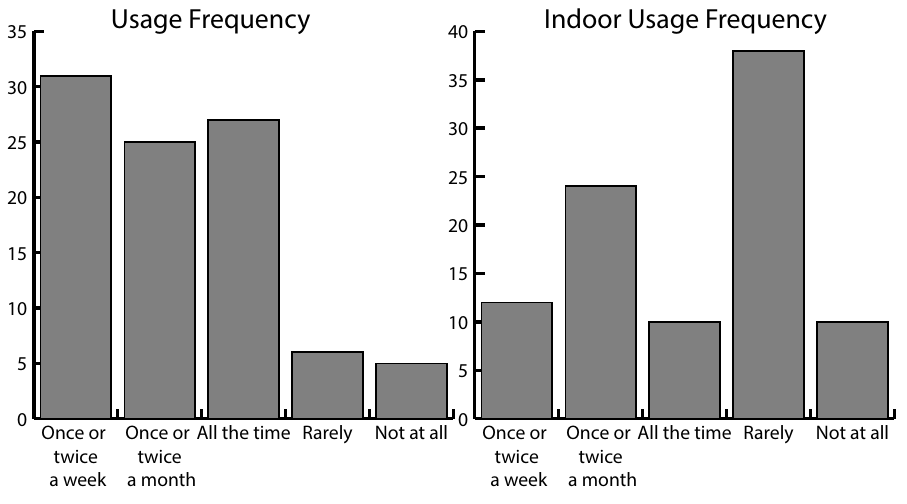}
	\caption{ \label{fig:frequency} Participants were asked to report how often they use their camera glasses normally and indoors}
\end{figure}

About one in three participants reported using their camera glasses once or twice a week, and one in four reported using them all the time. As expected for a device with sunglasses-style lenses, usage is primarily outdoors (see Figure \ref{fig:frequency}).

\subsection{Scenarios: How are camera glasses used?}

\begin{figure}
	\includegraphics[width=\linewidth]{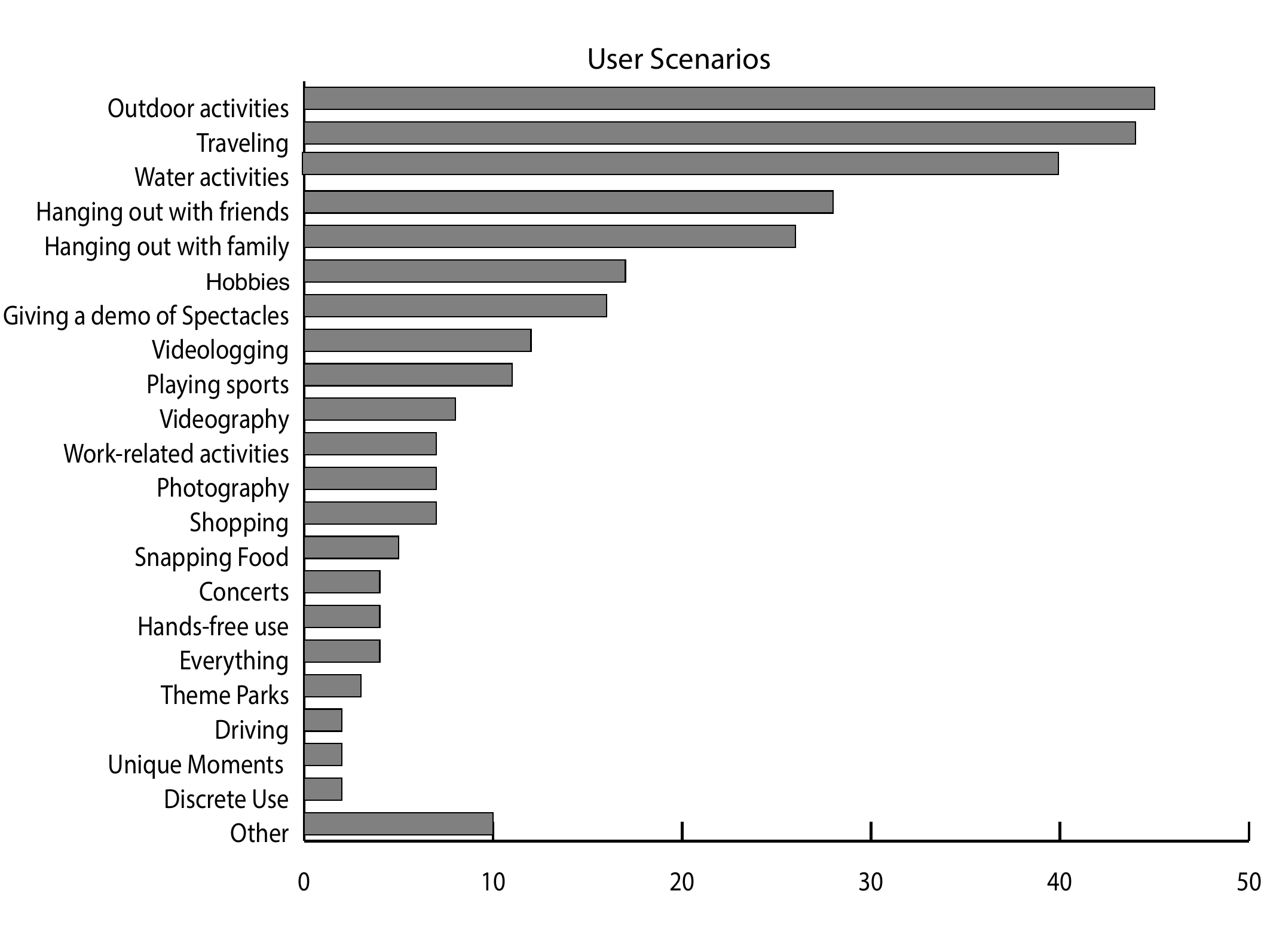}
    \caption{\label{fig:scenario} Most popular usage scenarios}
\end{figure}

Our surveys showed that camera glasses afford many different uses. We focus on the top four usage scenarios from the survey (see Figure ~\ref{fig:scenario}), which we also corroborated in the semi-structured interviews. 

Spectacles, the camera glasses investigated in our study, were created as a sunglasses-only product. User scenarios such as outdoor activities, traveling, and water activities are all situations reflecting this design choice. Approximately 23\% of survey respondents mentioned they would use Spectacles more if they were in more situations where sunglasses are appropriate. Furthermore, activities such as (indoor) concerts, shopping, and work-related activities showed less usage because people are less prone to wear sunglasses in these situations. 

Additionally, activities such as hanging out with friends and family were popular. This may be because users are more likely to use camera glasses in social situations to capture their memories. Camera glasses were less popular in more solitary activities. 
 
 \makeatletter
\renewenvironment{description}%
               {\list{}{\leftmargin=1pt
                        \labelwidth\z@ \itemindent-\leftmargin
                        \let\makelabel\descriptionlabel}}%
               {\endlist}
\makeatother

\begin{description}
\item[Outdoor activities.] People used their camera glasses for outdoor activities because it allowed them to capture an immersive and personal view of their environment: 
\begin{quote}\textit{
``I took them hiking and you could hear my breathing getting heavier and heavier as we climbed. So when you look back at it, you can see what I was seeing and feeling.''
 }---P24\end{quote}
 
Furthermore, having the camera glasses allowed participants to document collective outdoor activities for others:

\begin{quote}\textit{
``I took them skiing because I was with my family, girlfriend, younger brother. They were quite keen to get in the shots. They would all want some clips, while I was following them.''
 }---P12\end{quote}

\item[Traveling.]
The second most common use of the camera glasses was while traveling, especially when being ``out and about'':
 
\begin{quote}\textit{
``I use it for beach, travel... when I am going away, I can take pictures and videos of my destination. Those are really the big ones, when I am out and about exploring.''
 }---P11\end{quote}

Participants noted that camera glasses allowed them to capture their traveling adventures and share those experiences with others:
\begin{quote}\textit{
``I use it for travel and sightseeing. Since I live so close to Washington, I use it for travel. I get to bring people to other places.''
 }---P2\end{quote}

\item[Water activities.]
The release of the water-resistant feature in the second version of the Spectacles allowed people to capture water activities. These activities included trips to the beach and the pool, and having smooth transitions to and from non-water-related activities:
 
\begin{quote}\textit{
``We were playing around with a football, trying to throw a ball into the pool. We were recording with a camera, dunking into the pool.''
 }---P27\end{quote}

\item[Spending time with friends \& family]
People often cap\-tu\-red moments while hanging out with other people:
\begin{quote}
\textit{
``I use it mainly when I am with my friends. If I want to capture a moment, now I can get them together and take a picture.''
 }---P13\end{quote}
 
 Also, some people reporting letting others wear the glasses to allow them to capture their own perspective. For instance, a parent might let their child wear the glasses and capture their perspective.

\end{description}

 \subsection{Why are camera glasses used?}
 In our interviews, we were able to get a deeper understanding of why camera glasses are particularly fit for these circumstances. To help explain why camera glasses are used, we conducted a thematic analysis on the interview data. 
 Below, we report user quotes that describe the majority of reasons users felt that camera glasses were the best way to capture that moment. 

Our data show that capturing behavior using camera glasses is dependent on lifestyle factors. These factors include other devices people have, such as cellphones and action cameras, what types of careers they have, and what their daily lives look like. Another factor is what users do with content after capturing. In the discussion section, we extend these findings by postulating some factors that influence the reasons camera glasses are used. 

\begin{description}
\item[Capturing their own point of view.]
The development of new wearables has opened up the world of storytelling \cite{Rieger2018}, where users can capture content to present their own perspective on an event. As an example, participant P11 mentioned using the content captured with their camera glasses to view their own first-hand experiences:

\begin{quote}
\textit{
``My birthday, this time, last year, at a pool club around the city. We got a daybed with friends. Some of it I don't remember taking. Now I look back and it's fun to have the POV [point of view] shot.''
 }\---P11\end{quote}
 
Furthermore, camera glasses let users view their common activities from a different perspective: 
 
\begin{quote}\textit{
``I use them on days off. For sports, hanging out with friends, walking around, getting glimpses of my perspective.''
 }---P30\end{quote}

\item[Capturing special events.]
By understanding users' individual accounts of their usage, we learned about special events that camera glasses are used for. Most participants noted that they did not use their camera glasses for mundane activities, but rather for special ones.
Spectacles were created and designed to be integrated into users' lives as a pair of sunglasses. However, many users waited to capture special events where sunglasses could be worn, such as visiting a theme park or going to the vet: 
 
\begin{quote}
\textit{
``Yesterday, we went to Six Flags and we were wearing them on the roller coasters that aren't too extreme.''
 }---P1\end{quote}

\begin{quote}\textit{
``I adopted a cat. A little after I got the Spectacles, there was a yearly check up after I got her. This was the first time I was taking her out of my apartment. So I did some clips when I was driving, putting her in the carrier, taking her into the vet.''
 }---P11\end{quote}

\item[Capturing as an extension of daily life.]
Spectacles are not used all day, every day. However, at times they are used to augment daily activities. Some participants did mention integrating their camera glasses into their lives for more ``mundane'' activities, especially around other people:
 
\begin{quote}\textit{
``I usually use it when we go out with friends and are just walking around outside.''
 }---P5\end{quote}
  
\begin{quote}\textit{
``I use it when going to playground with my son, swimming, I have been to pool with them. One video was me taking it to the pool and you can see the water rushing towards you and it was great.''
 }---P24\end{quote}
 
\item[Capturing for work.]
Some people found professional uses for their camera glasses. For example, a participant who organizes school trips found camera glasses useful for documenting the trips:

\begin{quote}\textit{
``I go on field trips with my students and I have to take content for the weekly newsletter so this is the easiest way to do it without getting distracted.''
 }---P29\end{quote}
 
Similarly, a professional drone pilot used them to capture another perspective of flying a drone:
\begin{quote}\textit{
``They offer footage from multiple angles. I use them to show videos of me flying the drone, setting it up, etc... then upload videos and share them on Snapchat and Instagram.''
 }---P1\end{quote}

Participants in the entertainment and social media industry were able to capture scenes for their followers that provided novel, behind-the-scenes perspectives:
 
\begin{quote}\textit{
``It is a different way to show my followers what I am doing.''
 }---P10\end{quote}
 
\begin{quote}\textit{
``I capture behind the scenes things where I can capture the reaction of the thing then morph all the content into a video. }---P9\end{quote}

\item[Supplementing or replacing other capturing]
\item[technology.]
In the interviews, many participants noted that their camera glasses were a unique device that augmented or replaced other devices: 
 \begin{quote}\textit{
``I was going to use them in conjunction with my GoPro. Primarily POV, especially when I don't have my DSLR.''}---P21\end{quote}

Additionally, users even replaced their other capturing devices with camera glasses to capture their active lifestyles: 

\begin{quote}\textit{
``This was an interesting way to get GoPro content without having to strap it onto my body.''}---P32\end{quote}

Moreover, users were able to capture moments that would have been difficult to capture with other, more conspicuous technologies:

\begin{quote}\textit{
``Playing with my niece, I never imagined I could get her to take a picture. She threw a water balloon at me and it smashed me in the face. It was a neat perspective. Kids kind of freeze up when they see a camera, so I get a natural moment.''}---P29\end{quote}

Lastly, camera glasses allowed for hands-free capturing: over 25\% of survey participants mentioned purchasing Spectacles to capture hands-free content. From our interviews, this was one of the major reasons for capturing. It made capturing more accessible than other technologies in situations where people's hands and minds are occupied:
 
\begin{quote}\textit{
``One time I used it to capture my driving experiences, test driving, press the 10 second record, which would be hard to do it with my phone while driving. I was going from 0-60 mph, people got to see what I was seeing.''}---P15\end{quote}
\end{description}

\begin{description}
 \item[Capturing to share.]
Participants were often using their camera glasses to share content with their friends and family:

\begin{quote}\textit{
``I brought them during a trip to Italy, it was really nice because I came back and showed the content to my grandparents and it was like they went on the trip with me.''}---P2\end{quote}

Camera glasses also allowed people to share content from a specific event more easily. A participant mentioned being able to use their camera glasses to hold on to a memory and share it with other people attending the same event:

\begin{quote}\textit{
``Independence day, I had a lot of friends coming around. We were having BBQ, drinking beer, and everyone had a great time, and making sure the kids were alright. When they left, I shared the videos with everyone and they were shocked. It was a short memorable video for them and they really loved it.''}---P22\end{quote}

Sharing the device itself came up a few times. For instance one of the participants mentioned passing around their camera glasses during a particular event, so that at the end they could put together their own perspectives and share the entirety of the event:
 
\begin{quote}\textit{
``Well, I was hanging out with some of my friends, we started passing them around. People were sharing them, cool to see what people are looking at rather than a camera view.''}---P4\end{quote}

\item[Capturing to archive.]
Our findings show that users capture their own point of view with camera glasses to retain memorable events and share them with others. Prior work has shown that while social media are mostly about building strong social networks, they also have the potential of becoming a personal archive over time \cite{Zhao:2014:CTU:2556288.2557291}. We found in our interviews that camera glasses are used to capture and preserve memories. They have become a tool that aids in digitally archiving those memories for personal consumption: 

\begin{quote}\textit{
``I don't post everything, most of the time I just post it on my Snapchat, Sometimes I just want to record stuff for me''}---P14\end{quote}
\end{description}

\section{Discussion}
Our results helped us understand three key phenomena. First, people use camera glasses to capture, archive, and share with others. Second, these behaviors are influenced by societal norms that are still evolving because camera glasses adoption is still in its early stages. Third, people's behaviors are influenced by individual preferences such as lifestyle and aesthetics. These findings can be extended to other research on camera glasses, and open a research toolbox for designers to further study this space, which can ultimately lead to better camera glasses.

\subsection{Social interaction and privacy considerations}
Camera glasses still make up a small fraction of the wearables market \cite{wearables_consumers}. Their newness leads to curious reactions from those in the immediate environment of the user, because many people are not aware of the capabilities of the device. One participant noted that people asked them about it:
\begin{quote}\textit{
``When they talk to me, they often have this reaction of `Are you filming me now?'''} ---P30\end{quote}
The increased popularity of wearables has heightened designers' interest in the technology's privacy considerations. The designers of Spectacles approached this by adding a ring of LEDs that light up while recording. In our interviews, participants mentioned that people around them noticed the LEDs and often asked what they meant:
\begin{quote}\textit{
``I get a lot of `What is the light on the glasses?,' [in the] majority of the clips of people [they] are saying `What's that light?' [I tell them,] `They're Snapchat glasses'.''} ---P18\end{quote}
As more people gain awareness of devices like these, one might see changes in people's understanding of features like the LED ring and the sociocultural norms around the technology.
As with other types of cameras, users of camera glasses consider the privacy for themselves and the people around them. Prior research on lifeloggers has shown that these considerations occur at the moment of capturing content, like the participant above (P18) described \cite{Hoyle:2014:PBL:2632048.2632079}. Additionally, some of our participants mentioned thinking about privacy \emph{before} capturing too, and making a conscious decision about when it was appropriate to record:
\begin{quote}\textit{
`` ...respect people's privacy: the first thing I do is that. When I am taking a video, I don't use this for a spy game.''} ---P23\end{quote}
All of this highlights the value for researchers to ask how the design and use of camera glasses impact the privacy of both their users and the people around them.

\subsection{Individual preferences}
Camera glasses have the ability to take both videos and photos. However, users mentioned that the design of the device as glasses was better suited for capturing videos in active scenarios. This is further supported by the high number of users that use camera glasses for outdoor activities and traveling. Participants noted that video recording with the camera glasses allowed them to stay in the moment.
\begin{quote}\textit{
``The quality of me just moving around with glasses, recording is a lot better effect, the fish eye effect is really something.''} ---P2\end{quote}
\begin{quote}\textit{
``I think with photo, you try to take a perfect shot. I think in a short video, you actually manage to capture a life.''} ---P30 \end{quote}
The hands-free nature of capturing and novel features such as the HD circular video format influence what situations and types of content are better suited for use of the camera glasses. This opens up the door for new research on how features drive behavior and how we can continue to design to motivate effective usage.
Along with individual preferences in design and functionality, it was noted that users' lifestyle impacted their usage. Although some participants have been able to integrate their camera glasses into their daily activities, some consider their everyday activities ``too mundane'' and therefore not worth capturing. In the surveys, over 20\% of users mentioned that they would use Spectacles more if they discovered more opportunities for capturing. 
As two participants noted:
\begin{quote}\textit{
``Maybe it is because it was a new experience [at first], like I did not want to keep posting the same type of videos of me riding my bike everyday.''}---P23\end{quote}
\begin{quote}\textit{
``I wish I could record more if I did more stuff.''}---P5\end{quote}
Perhaps, like with the evolution from professional camera usage for special events to the use of camera phones for capturing everyday life, the culture around camera glasses may evolve to make it more common to use the technology for mundane activities.
It is important for researchers to continue to ask what individual lifestyle differences impact the post-adoption usage of camera glasses to further develop products that fit within these dependencies.

\section{Limitations}
While this paper focuses on laying out a typology of user behavior around camera glasses, we only interacted with people who used one of two versions of one particular product. More work is needed to compare usage of these devices, even across versions of the same product, and other products.

Furthermore, like in any study, the participants in our study may be different from the general population 
because of self-selection bias. Future work could focus on performing quantitative log data analyses on larger and more representative samples. 

\section{Conclusion and future work}
We classified the scenarios in which people use camera glasses, and what motivates them to capture content in these situations. Drawing from both surveys and semi-structured interviews with Spectacles users, we showed how societal and individual barriers influence usage.

Future research can extend this work by studying the content captured with camera glasses, i.e., rather than asking people what they do, analyze the traces of what they do. For instance, one could grab publicly available content captured with camera glasses and perform manual or automated analyses (e.g., using computer vision) to identify common scenes and scenarios where these devices are used. For this study, we did not use actual content captured by users, because it involves a trade-off between participant privacy and potential useful insights that could be elicited from the data. Assuming only a subset of participants would have given us approval to analyze their user content, we would have had little data to draw upon. From our perspective, this was not worth the privacy trade-off.

Lastly, as the technology evolves, researchers need to examine the social implications of camera glasses, and how devices with different aesthetics and features are adopted, and shape or get shaped by the culture around the people who use them. 

\begin{acks}
We thank the Spectacles team at Snap Inc. for their assistance with this project.
\end{acks}


\end{document}